# All-Fiber Source of Polarization-Entangled Photon Pairs Based on a Novel Birefringence Compensated Scheme

Han Chuen Lim, Dexiang Wang, Takuo Tanemura, Kazuhiro Katoh, and Kazuro Kikuchi
Research Center for Advanced Science and Technology, The University of Tokyo.
Room 512, Building 3, 4-6-1, Komaba, Meguro-ku, Tokyo 153-8904.
Email address: hanchuen@ginjo.rcast.u-tokyo.ac.jp

**Abstract:** We propose a new all-fiber source of polarization-entangled photon pairs for quantum communications. Fiber birefringence is compensated using Faraday rotator mirror, resulting in enhanced stability against random polarization drifts compared to existing schemes.

## 1. Introduction

Entangled-photons are necessary for many quantum communication protocols, such as quantum dense coding [1] and quantum teleportation [2]. To implement these protocols over optical fiber links, it is desirable to employ entangled photons having wavelengths in the 1550 nm telecom band, where loss of optical fiber can be as low as 0.2 dB/km. Among the various types of photon entanglement, polarization entanglement has attracted considerable attention in recent years mainly because of ease of polarization manipulation. While polarization-entangled photon pair production in the visible wavelength range via spontaneous parametric down conversion in nonlinear optical crystals was proposed and demonstrated roughly a decade ago [3,4], generating such pairs at the telecom band via parametric fluorescence using fiber nonlinearity was demonstrated only very recently [5-7]. Other non-fiber proposals at the telecom wavelength include the use of periodically-poled lithium niobate waveguides [8].

In the fiber proposals, the basic idea is to first separate a diagonally-polarized pump pulse into orthogonally-polarized components (or use two orthogonally-polarized pump pulses locked in phase), then inject them into an optical fiber so that they independently produce quantum-correlated photons via parametric fluorescence. Orthogonally-polarized photons are then combined at the output, erasing any information regarding their origin. Indistinguishability results in polarization entanglement. In [5,6], orthogonally-polarized, 5-ps pump pulses are first given a time delay of about 30 ps before input into optical fiber. After quantum correlated photons have been produced, a piece of 20-m-long polarization-maintaining fiber (PMF) is used to remove the time delay. A Sagnac interferometer configuration and a double-grating spectrometer are employed to achieve the necessary pump rejection ratio of larger than 100 dB. This configuration is rather difficult to implement in practice because due to the long fiber length, environment-induced birefringence changes occur in the fiber. This leads to random drifts of light polarization in the fiber, so adjustment of an in-loop polarization controller (PC) from time to time is required for effective pump suppression. In addition, adjustment of a second PC before the PMF is also necessary for exact compensation of the time delay. In [7], the use of a polarization-diversity loop allows orthogonally-polarized pump pulses to propagate in opposite directions in a fiber loop so that quantum-correlated photons automatically recombine upon leaving the setup. Although this scheme is simpler to implement, tuning of in-loop PCs is also necessary to compensate random polarization drifts that occur in the fiber loop. In this work, we propose a new all-fiber polarization-entangled photon pair source that incorporates fiber birefringence compensation so that the setup is insensitive to polarization drifts that occur in the long optical fiber used to produce parametric fluorescence.

## 2. Proposal

Figure 1 shows the principle of our proposed configuration. Similar to the polarization-diversity loop scheme [7], our scheme requires the input pump pulses be linearly-polarized 45 degrees with respect to the principal axes of a polarization beam-splitter (PBS). A 2-m-long polarization-maintaining fiber connecting two adjacent ports of the PBS serves to separate pump pulses into vertically- and horizontally-polarized components by a time delay of about 10 ns, as depicted in the figure. The resulting orthogonally-polarized pump pulses then independently produce quantum-correlated photons in an optical fiber. The Faraday rotator mirror (FRM) in double-pass configuration compensates the random birefringence of the optical fiber completely [9]. It is easy to see that the time-delay is automatically removed at the output port of the PBS and a circulator can then be used to extract the output light.

## 3. Experiment

Figure 2 shows our experimental setup. A continuous-wave laser of wavelength 1551.1 nm is intensity-modulated to produce pump pulses with a pulse width of 1 ns and repetition rate of 1 MHz. The pulses are then amplified using an erbium-doped fiber amplifier. We use a cascade of three band-pass optical filters to suppress amplified spontaneous emission noise. The pump pulses are then input into our proposed setup. The fiber that we use is a 1-km-long highly-nonlinear dispersion-shifted fiber having a nonlinear coefficient of 20 $W^{-1}km^{-1}$. We select the laser wavelength to be very close to the zero-dispersion wavelength of the fiber so as to ensure efficient generation of quantum-correlated photon pairs. We use four FBGs to suppress the pump by more than 100 dB. After pump rejection, we use two FBGs and two circulators to drop photons at wavelengths 1549.3 nm and 1552.9 nm for detection with commercially available single-photon detectors (Id Quantique). The gating width is 2.5 ns and the gating rate is 1 MHz. Dark counts are negligible. A 0.5-nm band-pass filter is placed before each detector to further remove residual pump photons. We measured the number of coincidence counts in 20 seconds for –5.5 dBm of input average power. Figure 3 shows the change in the coincidence-count rate (after subtracting accidental coincidences) observed when we rotated one of the half-wave-plates. Angle of the polarizer at the other channel was set to $-22.5^o$ (black circles), $22.5^o$ (white circles), $67.5^o$ (black triangles) and $112.5^o$ (white triangles). The result reveals a slight violation of the CHSH inequality (S = 2.35), which confirms that the output photons are entangled in polarization. Overall photon collection and detection efficiency of our detection setup was less than 1% for each channel, leaving room for improvement. Nevertheless, we can still observe polarization-entanglement from our proposed source. Accidental coincidences that are due to Raman scattering can be reduced by cooling the fiber [10].

## 4. Conclusion

We have proposed a novel all-fiber source of polarization-entangled photon pairs that is stable against polarization drifts in optical fiber. The scheme is easy to set up using commercially available fiber-optic components and is thus promising as a practical source of polarization-entangled photon pairs for quantum communications.

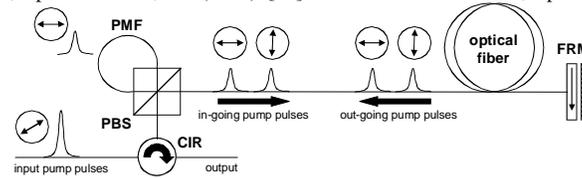

**Fig. 1**

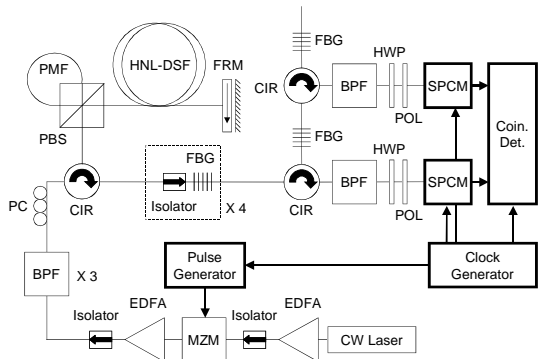

**Fig. 2**

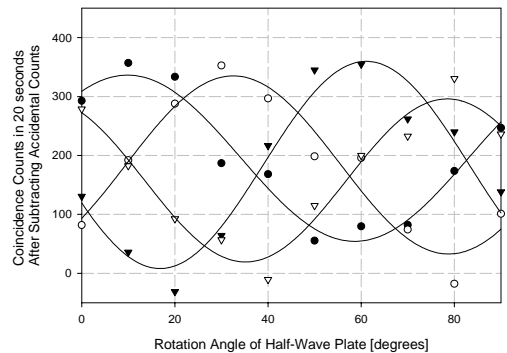

**Fig. 3**

Fig. 1 Proposed scheme. CIR: circulator, FRM: Faraday rotator mirror, PBS: polarization beam-splitter, PMF: polarization-maintaining fiber.
Fig. 2 Experimental setup. BPF: band-pass filter, Coin. Det.: coincidence detection, CW laser: continuous-wave laser, EDFA: erbium-doped fiber amplifier, FBG: fiber Bragg grating, HNL-DSF: highly-nonlinear dispersion-shifted fiber, HWP: half-wave plate, MZM: Mach-Zehnder modulator, PC: polarization controller, POL: polarizer, SPCM: single-photon detector module.
Fig.3 Change in number of coincidence counts in 20 seconds (minus accidental counts) with rotating angle of half-wave plate, for polarizer angles of $-22.5^o$ (black circles), $22.5^o$ (white circles), $67.5^o$ (black triangles) and $112.5^o$ (white triangles). The lines are sinusoidal fits.